\documentclass{article}
\usepackage{spconf,amsmath,graphicx}

\usepackage[utf8]{inputenc} 
\usepackage[T1]{fontenc}    
\usepackage{hyperref}       
\usepackage{url}            
\usepackage{booktabs}       
\usepackage{amsfonts}       
\usepackage{nicefrac}       
\usepackage{microtype}      
\usepackage{xcolor}         
\usepackage{cite}
\usepackage{float}

\usepackage{bm}
\usepackage{tikz}
\usepackage{pgfplots}
\usepackage{subcaption}
\usepackage{glossaries}
\usepackage{placeins}
\usepackage{array}
\usepackage{multirow}

\pgfplotsset{compat=1.16}
\usetikzlibrary{pgfplots.colormaps}
\usetikzlibrary{arrows.meta}
\usetikzlibrary{3d,arrows,automata,backgrounds,calc,calendar,chains,decorations,decorations.footprints,decorations.fractals,decorations.markings,decorations.pathmorphing,decorations.pathreplacing,decorations.shapes,decorations.text,er,fadings,fit,folding,matrix,mindmap,patterns,petri,plothandlers,plotmarks,positioning,scopes,shadows,shapes.arrows,shapes.callouts,shapes,shapes.gates.logic.IEC,shapes.gates.logic.US,shapes.geometric,shapes.misc,shapes.multipart,shapes.symbols,through,topaths,trees,pgfplots.groupplots}
\usepackage{tkz-kiviat-new}
\usepgfplotslibrary{statistics}

\title{Test-Time Adaptation for Speech Enhancement via Domain Invariant Embedding Transformation}
\name{Tobias Raichle, Niels Edinger, Bin Yang}
\address{Institute of Signal Processing and System Theory \\ University of Stuttgart}

\glsdisablehyper
\newacronym{rir}{RIR}{room impulse response}
\newacronym{se}{SE}{speech enhancement}
\newacronym{tta}{TTA}{test-time adaptation}
\newacronym{ttt}{TTT}{test-time training}
\newacronym{uda}{UDA}{unsupervised domain adaptation}
\newacronym{asr}{ASR}{automatic speech recognition}

\newacronym{ssra}{SSRA}{self-supvervised representation based adaptation}
\newacronym{pfpl}{PFPL}{phone-fortified perceptual-loss}
\newacronym{laden}{LaDen}{latent denoising}
\newacronym{diet}{DIET}{domain invariant embedding transformation}

\newacronym{ss}{SS}{spectral subtraction}
\newacronym{am}{AM}{amplitude masking}

\newacronym{mse}{MSE}{mean squared error}
\newacronym{pesq}{PESQ}{perceptual evaluation of speech quality}
\newacronym{sisdr}{SI-SDR}{scale-invariant signal-to-distortion ratio}
\glsunset{sisdr}
\newacronym{ssnr}{SSNR}{segmental signal-to-noise ratio}
\glsunset{ssnr}

\newacronym{vbd}{VBD}{VoiceBank+DEMAND}
\newacronym{vbw}{VBW}{VoiceBank+WHAM!}
\newacronym{dns}{DNS}{deep noise suppression}

\begin{document}
\definecolor{mittelblau}{RGB}{0, 126, 198}
\definecolor{violettblau}{cmyk}{0.9, 0.6, 0, 0}
\definecolor{rot}{RGB}{238, 28 35}
\definecolor{apfelgruen}{RGB}{140, 198, 62}
\definecolor{gelb}{RGB}{255, 229, 0}
\definecolor{orange}{RGB}{244, 111, 33}
\definecolor{pink}{RGB}{237, 0, 140}
\definecolor{lila}{RGB}{128, 10, 145}
\definecolor{cyan}{RGB}{58, 252, 252}
\definecolor{moosgruen}{RGB}{53, 147, 134}
\definecolor{bordeaux}{RGB}{148, 23, 112}
\definecolor{kastanie}{RGB}{148, 23, 81}
\definecolor{lachs}{RGB}{255, 126, 121}
\definecolor{turkis}{HTML}{B3ECEC}
\definecolor{lavendel}{RGB}{215, 131, 254}
\definecolor{hellgrau}{RGB}{224, 224, 224}
\definecolor{mittelgrau}{RGB}{128, 128, 128}
\definecolor{dunkelgrau}{RGB}{80,80,80}
\definecolor{anthrazit}{RGB}{19, 31, 31}

\pgfplotsset{bar_1_style/.style={color=mittelblau, fill=mittelblau!50}}
\pgfplotsset{bar_2_style/.style={color=orange, fill=orange!50}}
\pgfplotsset{bar_3_style/.style={color=pink, fill=pink!50}}
\pgfplotsset{bar_4_style/.style={color=lila, fill=lila!50}}
\pgfplotsset{bar_5_style/.style={color=lavendel, fill=lavendel!50}}

\pgfplotscreateplotcyclelist{bar_default}{%
bar_1_style\\%
bar_2_style\\%
bar_3_style\\%
bar_4_style\\%
bar_5_style\\%
}

\newcommand{\PreserveBackslash}[1]{\let\temp=\\#1\let\\=\temp}
\newcommand{\newcol}{\vfill\pagebreak}

\newcolumntype{C}[1]{>{\PreserveBackslash\centering}p{#1}}

\tikzset{arrow/.style={
->,
>={Stealth[inset=0pt, angle=30:5pt]},
}}

\tikzset{arrow_large/.style={
->,
>={Stealth[inset=0pt, angle=30:12pt]},
}}

\tikzset{line_back/.style={
			color=lila,
		}}

\tikzset{arrow_back/.style={
<-,
>={Stealth[inset=0pt, angle=35:4pt]},
color=lila,
}}

\newlength{\wcol}

\newcommand{\matr}[1]{\mathbf{#1}}
\newcommand{\vect}[1]{\bm{#1}}

\newcommand{\cmark}{\ding{51}}%
\newcommand{\xmark}{\ding{55}}%
\newcommand{\glsi}[1]{\glsunset{#1}\gls{#1} (\acrlong{#1})}
\newcommand{\Glsi}[1]{\glsunset{#1}\Gls{#1} (\acrlong{#1})}

\newcommand{\dns}[1]{\(\text{DNS}_{\text{#1}}\)}
\newcommand{\avg}[1]{\(\overline{\text{#1}}\)}

\newcommand{\todo}[1]{{\color{red}TODO: }#1}

\newlength{\corner}
\newlength{\scorner}

\tikzstyle{frozen} = [double, thick, draw=turkis, fill=turkis!50, text=turkis!330, rounded corners=\scorner]
\tikzstyle{truth} = [draw=apfelgruen, fill=apfelgruen!10, text=apfelgruen!120, rounded corners=\scorner]
\tikzstyle{trainable} = [draw=lila, fill=lila!10, text=lila!80!black, semithick, rounded corners=\scorner]
\tikzstyle{norm} = [draw=orange, fill=orange!10, text=orange!90!black, semithick]
\tikzstyle{signal} = [draw=mittelgrau, text=mittelgrau!90!black, semithick, rounded corners=\corner, inner sep=0pt]
\tikzstyle{dsp} = [draw=moosgruen, fill=moosgruen!20, text=moosgruen!90!black, semithick, rounded corners=\scorner]

\pgfdeclarepatternformonly{south west lines}{\pgfqpoint{-0pt}{-0pt}}{\pgfqpoint{3pt}{3pt}}{\pgfqpoint{3pt}{3pt}}{
	\pgfsetlinewidth{0.4pt}
	\pgfpathmoveto{\pgfqpoint{0pt}{0pt}}
	\pgfpathlineto{\pgfqpoint{3pt}{3pt}}
	\pgfpathmoveto{\pgfqpoint{2.8pt}{-.2pt}}
	\pgfpathlineto{\pgfqpoint{3.2pt}{.2pt}}
	\pgfpathmoveto{\pgfqpoint{-.2pt}{2.8pt}}
	\pgfpathlineto{\pgfqpoint{.2pt}{3.2pt}}
	\pgfusepath{stroke}}

\pgfdeclarepatternformonly{south east lines}{\pgfqpoint{-0pt}{-0pt}}{\pgfqpoint{3pt}{3pt}}{\pgfqpoint{3pt}{3pt}}{
	\pgfsetlinewidth{0.4pt}
	\pgfpathmoveto{\pgfqpoint{0pt}{3pt}}
	\pgfpathlineto{\pgfqpoint{3pt}{0pt}}
	\pgfpathmoveto{\pgfqpoint{.2pt}{-.2pt}}
	\pgfpathlineto{\pgfqpoint{-.2pt}{.2pt}}
	\pgfpathmoveto{\pgfqpoint{3.2pt}{2.8pt}}
	\pgfpathlineto{\pgfqpoint{2.8pt}{3.2pt}}
	\pgfusepath{stroke}}


%
\maketitle
\begin{abstract}
	Deep learning-based speech enhancement models achieve remarkable performance when test distributions match training conditions, but often degrade when deployed in unpredictable real-world environments with domain shifts.
	To address this challenge, we present \glsi{laden}, the first test-time adaptation method specifically designed for speech enhancement.
	Our approach leverages powerful pre-trained speech representations to perform latent denoising, approximating clean speech representations through a linear transformation of noisy embeddings.
	We show that this transformation generalizes well across domains, enabling effective pseudo-labeling for target domains without labeled target data.
	The resulting pseudo-labels enable effective test-time adaptation of speech enhancement models across diverse acoustic environments.
	We propose a comprehensive benchmark spanning multiple datasets with various domain shifts, including changes in noise types, speaker characteristics, and languages.
	Our extensive experiments demonstrate that LaDen consistently outperforms baseline methods across perceptual metrics, particularly for speaker and language domain shifts.
\end{abstract}

\glsresetall
\section{INTRODUCTION}
\subsection{Motivation}
Modern deep learning based \gls{se} models have achieved remarkable success and are able to deliver natural sounding denoised speech even for very noisy recordings.
However, previous works focus mostly on the performance on a small number of benchmark datasets, e.g. the ubiquitous \gls{vbd}~\cite{vbd} dataset, whereas generalization has not received as much attention.
This has resulted in \gls{se} models performing exceptionally well, as long as the target data distribution closely matches the training distribution, but diminished performance under distribution shifts.
Table~\ref{tab:pre_comp_cmgan} shows that the \gls{se} model CMGAN~\cite{cmgan} trained on the source EARS-W~\cite{ears} dataset performs significantly worse on the target datasets (EARS-D and VBD) than the same model trained directly on those target domains.
Since \gls{se} models are commonly deployed in unpredictable environments, generalization is a key factor for successful, practical speech enhancement systems.
Due to this unpredictability, training \gls{se} models on all possible target distributions is not feasible.
To nonetheless achieve generalization, models must be able to adapt themselves to any test environment without relying on labeled data from the specific target domain.
Methods that perform adaptation under these conditions can be summarized under \gls{uda}.
\par
In practice, adaptation methods cannot rely on the availability of labeled source data at test time.
Besides the privacy concerns of sharing labeled source data, it is not practical to store and re-process the large source domain dataset during adaptation~\cite{tent}.
This is especially true for speech enhancement, since these models are commonly deployed to edge devices with limited storage and compute resources.
Additionally, real-world deployments demand adaptation to happen simultaneously to inference, as separate adaptation phases would introduce unacceptable latency.
Online adaptation using only the source model and unlabeled target data is known as \gls{tta} and presents the most general and practical adaptation paradigm \cite{roid}.
While there are previous works covering \gls{uda} in the context of speech enhancement, to the best of our knowledge, this is the first work exploring \gls{tta} for speech enhancement.
\par
The main contributions of this work can be summarized as follows:
\begin{itemize}
	\item We explore the application of \gls{tta} to speech enhancement and propose a comprehensive benchmark for evaluation, spanning various domain shifts, including different noise types, speaker characteristics, and languages.
	\item We propose and empirically verify \glsi{diet} as an effective method for translating between noisy and clean speech in embedding spaces.
	\item Based on \gls{diet} we introduce LaDen (latent denoising), a novel approach that enables \gls{tta} for speech enhancement.
	\item By conducting a large number of experiments, we identify the strengths and limitations of the proposed method across different domain shifts.
\end{itemize}

\begin{table}[h]
	\def\arraystretch{1}
	\centering
	\caption{Comparison between the source CMGAN trained on the EARS-W~\cite{ears} dataset and models trained on the respective target domain (target CMGAN). EARS-D denotes the dataset introduced in Section~\ref{sec:exp}. SI-SDR in dB.}
	\label{tab:pre_comp_cmgan}
	\small\setlength{\tabcolsep}{5pt}
	\begin{tabular}{lcccc}
	\toprule
	             & \multicolumn{2}{c}{EARS-D} & \multicolumn{2}{c}{VBD}                                           \\
	\cmidrule(r){2-3} \cmidrule(r){4-5}
	             & PESQ \(\uparrow\)          & SI-SDR  \(\uparrow\)    & PESQ \(\uparrow\) & SI-SDR \(\uparrow\) \\
	\midrule
	No denoising & 1.398                      & -1.109                  & 1.970             & 8.444               \\
	Source CMGAN & 2.701                      & 4.091                   & 3.234             & 10.094              \\
	Target CMGAN & \textbf{2.763}             & \textbf{11.823}         & \textbf{3.399}    & \textbf{20.071}     \\
	\bottomrule
\end{tabular}

\end{table}

\subsection{Problem Setting}
Given a recording of corrupted speech \(\vect{y}\in \mathcal{A}\), the goal of \gls{se} is to estimate the uncorrupted speech \(\vect{x} \in \mathcal{A}\), where \(\mathcal{A}\) denotes the set of all audio signals~\cite{speech_enhancement}.
Generally, corruptions can include additive noise, reverberation and echoes, limited bandwidth or compression artifacts.
While the proposed method can be applied to all distortion types, this work only considers additive noise and leaves other corruptions for future research.
Thus, the setting can be modeled by
\begin{align*}
	\vect{y}=\vect{x}+\vect{n},
\end{align*}
where \(\vect{n} \in \mathcal{A}\) denotes the additive noise.
The task of the \gls{se} model \({f_{\theta}:\mathcal{A}\rightarrow\mathcal{A}}\) is to estimate the clean signal \(\hat{\vect{x}}\in\mathcal{A}\)
\begin{align*}
	\hat{\vect{x}}=f_{\theta}(\vect{y}).
\end{align*}
The source \gls{se} model \(f_{\theta}\) is trained using the labeled source dataset \({\mathcal{D}_{\mathrm{S}}\subset \mathcal{A}\times\mathcal{A}}\), consisting of pairs of clean and noisy speech segments.
\par
Given the trained source model, the task of \gls{tta} is to adapt the model with only the unlabeled target dataset \({\mathcal{D}_{\mathrm{T}}\subset \mathcal{A}}\), while simultaneously performing inference.
We assume a distribution shift between the source (\(\mathrm{S}\)) and target (\(\mathrm{T}\)) datasets that can manifest itself in a shift in the speech distribution \(p_{X}\) and/or the noise distribution \(p_{N}\), resulting in the mismatch \(p_{\mathrm{S};X,N}\neq p_{\mathrm{T};X,N}\) and therefore \(p_{\mathrm{S};Y}\neq p_{\mathrm{T};Y}\).
However, the predictive distribution \(p_{X|Y}\), i.e., the \gls{se} task, stays the same \(p_{\mathrm{S};X|Y}(\vect{x}|\vect{y})=p_{\mathrm{T};X|Y}(\vect{x}|\vect{y})\) \cite{ssa,comet}.
\Gls{tta} assumes the model \(f_{\theta}\) as given, i.e. no changes to the source training or model architecture can be made.
This is in contrast to \gls{ttt}, where an auxiliary self-supervised task is added during source training to be used later for adaptation~\cite{ttt}.
\par
In the field of \gls{tta} concerning classification, most methods rely on the probabilistic model output to perform adaptation~\cite{ssra}.
The main approaches can be summarized under entropy minimization~\cite{tent}, feature alignment~\cite{ssa,bn_tta} and pseudo-labeling~\cite{rmt}, each with various extensions to improve stability~\cite{eata,roid}, generalization~\cite{rmt,deyo} or efficiency~\cite{cotta}.
Clearly, entropy minimization does not easily translate to \gls{se} because \gls{se} models output a direct estimate of the clean signal instead of a probability distribution.
Feature alignment typically enforces consistency of model features under label-preserving input perturbations~\cite{tta_feat}.
However, as \gls{se} models modify their input, they cannot be invariant to augmentations that affect the speech component.
Designing an output preserving augmentation is therefore non-trivial, as the assumption of general invariance to small perturbations is not valid.
Further, designing a consistency metric that is well aligned with the perceptual \gls{se} task poses an additional challenge.
A subset of feature alignment methods update the model's batch normalization statistics~\cite{bn_tta}, but this has limited applicability in \gls{se}, where many architectures (including those used in this study) do not rely on batch normalization layers.
Similarly to entropy minimization, estimating pseudo-labels, i.e., computing a proxy for the clean signal, is not straightforward for \gls{se}, as the direct signal estimation in \gls{se} as a regression task does not offer a comparable way of assigning an estimated label in classification.
An effective method for computing pseudo-labels thus remains a key challenge in achieving \gls{tta} for \gls{se}.

\section{RELATED WORK}
\label{sec:related}
As \gls{se} methods commonly experience a domain shift in practical deployments, applying \gls{uda} to \gls{se} has been an active field of research.
In \cite{dotn}, the authors propose phrasing the domain adaptation problem as an optimal transport problem.
Put simply, given a target sample, the most similar noisy source sample is identified and the corresponding clean source sample is used as a pseudo-label to perform adaptation.
Since this approach assumes access to the source dataset, it is not compatible with the \gls{tta} setting.
\cite{msp} leverages a two-stage approach, that, similarly to \gls{ttt}~\cite{ttt}, uses masked spectrogram prediction as a self-supervised auxiliary task to adapt the model.
Not only does this approach also rely on source data, it is also inherently offline since the adaptation precedes the \gls{se} training.
RemixIT~\cite{remixit} leverages self-training using a student-teacher approach.
Given a set of noisy target samples, a trained teacher model estimates the clean speech and noise components.
These estimates are used to create a weakly labeled dataset by permuting the components to create new bootstrapped pairs of clean and noisy speech.
The \gls{se} student model is then adapted using this dataset.
This \gls{uda} approach does not rely on source data and therefore conforms to the source-free online \gls{tta} paradigm studied in this work, although it has never been explored in this setting.
It is therefore used as a baseline in this work.
The \gls{uda} method that is most closely related to this work is \glsi{ssra}~\cite{ssra}.
The approach is similar to \cite{dotn} in that the most similar source samples are identified and used as pseudo-labels.
However, \acrshort{ssra} identifies and compares the most similar source samples in the latent space spanned by wav2vec~\cite{wav2vec}.
It can be interpreted as a transfer of PFPL~\cite{pfpl} to the \gls{uda} paradigm.
\glsi{pfpl} proposes improving supervised \gls{se} training by minimizing the Wasserstein distance between clean and noisy embeddings.
In~\cite{tta_pers_se}, distilling a source-trained general \gls{se} teacher model into a smaller, personalized student model is phrased as a \gls{tta} problem.
However, while their approach conforms to the \gls{tta} paradigm, it does not address adapting beyond the generalization capability of the trained teacher.
\par
While these methods have advanced speech enhancement adaptation, they either rely on source data or operate offline, motivating our latent denoising approach that addresses these limitations.

\section{METHODOLOGY}
\label{sec:method}
\subsection{Domain Invariant Embedding Transformation}
The proposed method solves the problem of pseudo-labeling by constructing pseudo-labels in a semantic embedding space spanned by a speech encoder \(g:\mathcal{A}\rightarrow\mathbb{R}^{d}\) like wav2vec~\cite{wav2vec} or WavLM~\cite{wavlm} via a \acrfull{diet}.
The approach is based on the hypothesis that the relationship between noisy speech \(\vect{y}\) and clean speech \(\vect{x}\), which is highly complicated on the signal level, simplifies to a simple relationship between the noisy embeddings \(\vect{y}'=g(\vect{y})\) and the clean embeddings \(\vect{x}'=g(\vect{x})\).
Figure~\ref{fig:emb_transfer} illustrates this principle.
This implies that a simple model can be used to translate between noisy and clean embeddings.
For this work, a linear transformation is used to model this relationship in the embedding space by
\begin{align}
	\label{eq:approx}
	\vect{x}'\approx \matr{A} \vect{y}'
\end{align}
with \(\matr{A}\in\mathbb{R}^{d\times d}\).
Given a number of samples \(K\geq d\), the transformation can be estimated via
\begin{align}
	\label{eq:approx_fit}
	\matr{A}=\matr{X}'\matr{Y}'^{+},
\end{align}
where \(\matr{X}',\matr{Y}' \in \mathbb{R}^{d\times K}\) denote \(K\) stacked embedding vectors \(\vect{x}'\) and \(\vect{y}'\), respectively and \(\matr{Y}'^{+}\in \mathbb{R}^{K\times d}\) denotes the Moore-Penrose inverse of \(\matr{Y}'\).
\par
Crucially, our experiments show that the transformation \(\matr{A}\) generalizes across domains with surprising accuracy.
Hence, it is largely domain invariant.
Table~\ref{tab:mat_gen} shows the cosine similarity \({\operatorname{sim}(\vect{x}',\vect{y}')=\frac{\vect{x}'^{T}\vect{y}'}{\Vert \vect{x}' \Vert\cdot \Vert \vect{y}' \Vert}}\) between ground truth clean embeddings \(\vect{x}'\) and estimated clean embeddings \(\matr{A}\vect{y}'\) using the speech encoder \(g\) from~\cite{wavlm} and the \gls{diet} matrix \(\matr{A}\) fitted on the EARS-W dataset.
\begin{table*}[h]
	\def\arraystretch{1}
	\centering
	\caption{\Gls{diet} accuracy, i.e. the average cosine similarity between ground truth clean embeddings \(\vect{x}'\) and noisy embeddings \(\vect{y}'\) as well as transformed noisy embeddings \(\matr{A}\vect{y}'\) for different target datasets.
		\(\matr{A}\) was estimated using the EARS-W training split and is evaluated on the respective test split.
		The datasets are described in detail in Section~\ref{sec:exp}.
		Highest similarity is \textbf{bold}.}
	\label{tab:mat_gen}
	\begin{tabular}{lccccccccc}
	\toprule
	                                                    & EARS-W          & EARS-D          & VBD             & VBW             & \dns{EN}        \\
	\midrule
	\(\operatorname{sim}(\vect{x}',\vect{y}')\)         & 0.8618          & 0.9062          & 0.8857          & 0.7627          & 0.8765          \\
	\(\operatorname{sim}(\vect{x}',\matr{A}\vect{y}')\) & \textbf{0.9941} & \textbf{0.9927} & \textbf{0.9766} & \textbf{0.9727} & \textbf{0.9663} \\
	\bottomrule
\end{tabular}

\end{table*}
\par

\subsection{Latent Denoising}
\begin{figure}[h]
	\centering
	\begin{subfigure}[t]{0.45\textwidth}
		\resizebox{0.9\textwidth}{!}{\begin{tikzpicture}
	\begin{axis}[
			tick align=outside,
			tick pos=left,
			x grid style={black},
			xtick style={color=black},
			y grid style={darkgray176},
			ytick style={color=black},
			ymin=-1.4,
			height=5cm, width=10cm,
			legend style={font=\small, legend columns=2,at={(0.5,-0.2)}, anchor=center},
			legend cell align={left},
			hide axis,
		]
		\addplot [
			only marks,
			mark size=1.5,
			mark=triangle*,
			mittelblau,
			legend image post style={draw=gray, fill=gray, mark size=2.4pt},
		]
		table [x=x, y=y]
			{figs/data/cloud_a.txt};
		\addlegendentry{Noisy}

		\addplot [
			only marks,
			mark size=1.5,
			mark=*,
			legend image post style={draw=gray, fill=gray, mark size=2.0pt},
			mittelblau,
		]
		table [x=x, y=y]
			{figs/data/cloud_b.txt};

		\addlegendentry{Clean}
		\addplot [
			only marks,
			mark size=1.5,
			mark=triangle*,
			orange,
			forget plot,
		]
		table [x=x, y=y]
			{figs/data/cloud_c.txt};

		\addplot [
			only marks,
			mark size=1.5,
			mark=o,
			orange,
			forget plot,
		]
		table [x=x, y=y]
			{figs/data/cloud_d.txt};

		\addlegendimage{
			legend image code/.code={
					\draw[draw=mittelblau, fill=mittelblau] (0cm,-0.1cm) rectangle (0.5cm,0.1cm);
				}
		}
		\addlegendentry{Source}
		\addlegendimage{
			legend image code/.code={
					\draw[draw=orange, fill=orange] (0cm,-0.1cm) rectangle (0.5cm,0.1cm);
				}
		}
		\addlegendentry{Target}

		\draw[arrow] (axis cs: 0,0) -- (axis cs: 8,0) node [midway, above] {\(\mathbf{A}_{\mathrm{S}}\)};
		\draw[arrow] (axis cs: 0,6) -- (axis cs: 9,5.5) node [midway, above] {\(\mathbf{A}_{\mathrm{T}} \approx {\mathbf{A}_{\mathrm{S}}}\)};

	\end{axis}

\end{tikzpicture}}
		\caption{\Acrfull{diet}.}
		\label{fig:emb_transfer}
		\vspace*{1em}
	\end{subfigure}
	\begin{subfigure}[t]{0.45\textwidth}
		\resizebox{\textwidth}{!}{\begin{tikzpicture}
	\tikzstyle{model} = [rectangle, draw=black, inner sep=1em, align=center]
	\tikzstyle{helper} = [inner sep=3pt]

	\setlength \scorner {2pt}
	\setlength \corner {5pt}

	\node (n_in) {\(\vect{y}\)};

	\node (m_SE) [right=2em of n_in, trainable, minimum size=4em, align=center] {SE \\ \(f_{\theta}\)};
	\node (m_FM_1) [right=2em of m_SE, frozen, minimum width=6em, minimum height=4em, align=center] {Encoder \\ \(g\)};
	\node (m_FM_2) [below= of m_FM_1, frozen, minimum width=6em, minimum height=4em, align=center] {Encoder \\ \(g\)};

	\node (sim) [right=5em of $(m_FM_1.east)!0.5!(m_FM_2.east)$, rectangle, draw, minimum size=2em] {\(\mathcal{L}_{\mathrm{LD}}\)};

	\coordinate [at=($(m_FM_2.east)!0.5!(sim.west)$)] (mid_sim);

	\node (mult) [circle, draw, at=(mid_sim|-m_FM_2.center), font=\large, inner sep=0pt] {\(\times\)};
	\node (A) [below=0.75em of mult] {\(\mathbf{A}\)};

	\coordinate [at=($(n_in.east)!0.5!(m_SE.west)$)] (mid_in);

	\draw [arrow] (n_in) -- (m_SE);
	\draw [arrow] (m_SE) -- (m_FM_1) node [midway, above] {\(\hat{\vect{x}}\)};
	\draw [arrow] (m_FM_1) -- (sim|-m_FM_1) node [near start, above] {\(\hat{\vect{x}}'\)} -- (sim);
	\draw [arrow] (mid_in) -- (mid_in|-m_FM_2) -- (m_FM_2);
	\draw [arrow] (m_FM_2) -- (mult) node [midway, above] {\(\vect{y}'\)};
	\draw [arrow] (A) -- (mult);
	\draw [arrow] (mult) -- (sim|-mult) -- (sim);

	\node (h_l) [helper, at=(sim.north)] {};
	\node (h_me) [helper, at=(m_FM_1.east)] {};
	\node (h_mw) [helper, at=(m_FM_1.west)] {};
	\node (h_s) [helper, at=(m_SE.east)] {};

	\draw [lila] (h_l.west) -- (h_me.south-|h_l.west) -- (h_me.south);
	\draw [arrow, lila] (h_mw.south) -- (h_s.south);

	\node (legend_origin) [helper, below=8em of n_in] {};
	\node (l_1_1) [at=(legend_origin), helper] {};
	\node (l_1_2) [right=1.5em of l_1_1, helper] {};
	\node (l_2_1) [below=0.75em of l_1_1, helper] {};
	\node (l_2_2) [right=1.5em of l_2_1, helper] {};

    \draw [arrow] (l_1_1) -- (l_1_2);
    \draw [arrow, lila] (l_2_1) -- (l_2_2);

    \node (l_f) [right=0em of l_1_2, font=\small] {Forward};
    \node (l_b) [right=0em of l_2_2, font=\small] {Backward};
    \node [fit=(l_1_1)(l_b)(l_f), draw, inner sep=2pt] {};

\end{tikzpicture}}
		\caption{Schematic of the proposed method \acrshort{laden}.}
		\label{fig:laden_schem}
	\end{subfigure}
	\caption{Using latent pseudo-labels for \gls{tta}.}
	\label{fig:laden}
\end{figure}
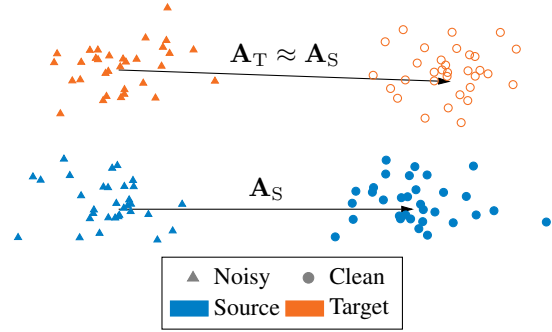
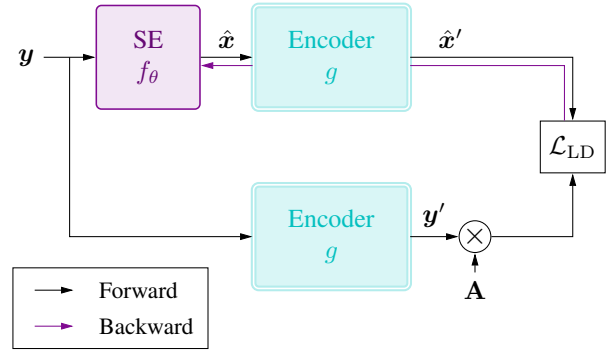

Based on \acrshort{diet}, \(\matr{A}\) can be estimated offline using the labeled source dataset \(\mathcal{D}_{\mathrm{S}}\), before using it online to compute pseudo-labels for the unlabeled target dataset \(\mathcal{D}_{\mathrm{T}}\).
This approach therefore does not violate the \gls{tta} paradigm.
Figure~\ref{fig:laden_schem} illustrates the principle of the adaptation approach LaDen.
The loss function \(\mathcal{L}_{\mathrm{LD}}\) is defined as the cosine distance, i.e., one minus the cosine similarity, between the pseudo-label \(\matr{A}\vect{y}'\) and the embedding of the \gls{se} output \(\hat{\vect{x}}'\)
\begin{align}
	\mathcal{L}_{\mathrm{LD}} = 1 - \operatorname{sim}(\hat{\vect{x}}',\matr{A}\vect{y}').
\end{align}
This self-supervised loss is then used to adapt the parameters of the \gls{se} model.

\par
For this work, we employ WavLM rather than other speech encoders like wav2vec, because WavLM's pre-training includes a denoising task.
This exposure to noisy speech during pre-training appears to create more robust and informative embeddings for both clean and noisy speech~\cite{superb}, enabling more accurate linear mapping in Eq.~\ref{eq:approx}.
Concretely, the CNN encoder of WavLM Large~\cite{wavlm} is used to generate the embeddings with a dimension of \(d=512\).
Using only the CNN encoder reduces the number of parameters from 316M to 4.2M.
Additionally, since the encoder's weights are frozen, the computational overhead is further reduced.
As WavLM splits the recording into frames and generates an embedding for each frame, the mean embedding of all frames per utterance is used.

\par
This latent denoising approach effectively addresses the pseudo-labeling challenge in test-time adaptation for speech enhancement by leveraging \gls{diet} to model the relationship between noisy and clean speech in the embedding space, while remaining computationally efficient.

\subsection{Envelope Regularization}
\begin{figure*}[h]
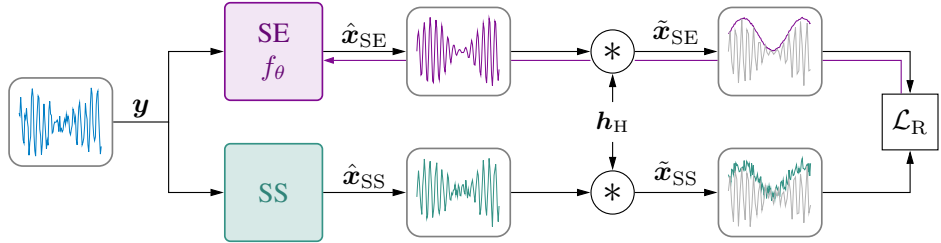

	\centering
	\def \npts {100}
	\resizebox{0.7\textwidth}{!}{\begin{tikzpicture}
	\tikzstyle{helper} = [inner sep=3pt]
	\setlength{\scorner}{2pt}
	\setlength{\corner}{5pt}

	\node (m_SE) [trainable, minimum size=3.5em, align=center] {SE \\ \(f_{\theta}\)};
	\node (m_SS) [dsp, below=1.5em of m_SE, minimum size=3.5em, align=center] {SS};

	\coordinate [at=($(m_SE.south west)!0.5!(m_SS.north west)$)] (mid_SE);

	\node (x_noisy) [left=4em of mid_SE] {\resizebox{3em}{!}{\input{figs/data/env_noisy}}};
	\node (x_noisy) [signal, at=(x_noisy), fit=(x_noisy)] {};
	\node (x_se) [right=3em of m_SE] {\resizebox{3em}{!}{\input{figs/data/env_se}}};
	\node (x_se) [signal, at=(x_se), fit=(x_se)] {};
	\node (x_ss) [right=3em of m_SS] {\resizebox{3em}{!}{\input{figs/data/env_ss}}};
	\node (x_ss) [signal, at=(x_ss), fit=(x_ss)] {};

	\node (m_SE_env) [draw, circle, right=of x_se, inner sep=2pt, align=center, font=\Large] {\(*\)};
	\node (m_SS_env) [draw, circle, right=of x_ss, inner sep=2pt, align=center, font=\Large] {\(*\)};

	\node (m_filt) [at=($(m_SE_env)!0.5!(m_SS_env)$), font=\small] {\(\vect{h}_{\mathrm{H}}\)};

	\node (x_se_env) [right=3em of m_SE_env] {\resizebox{3em}{!}{\input{figs/data/env_se_env}}};
	\node (x_se_env) [signal, at=(x_se_env), fit=(x_se_env)] {};
	\node (x_ss_env) [right=3em of m_SS_env] {\resizebox{3em}{!}{\input{figs/data/env_ss_env}}};
	\node (x_ss_env) [signal, at=(x_ss_env), fit=(x_ss_env)] {};

	\coordinate [at=($(x_se_env)!0.5!(x_ss_env)$)] (mid_env);

	\node (loss) [right=4em of mid_env, draw, minimum size=2em] {\(\mathcal{L}_{\mathrm{R}}\)};

	\coordinate [at=($(x_noisy.east)!0.5!(mid_SE)$)] (mid_SE);

	\draw [arrow] (x_noisy) -- node [midway, yshift=5pt] {\(\vect{y}\)} (mid_SE) -- (mid_SE|-m_SE) -- (m_SE);
	\draw [arrow] (mid_SE) -- (mid_SE|-m_SS) -- (m_SS);

	\draw [arrow] (m_SE) -- (x_se) node [midway, yshift=6pt] {\(\hat{\vect{x}}_{\mathrm{SE}}\)};
	\draw [arrow] (m_SS) -- (x_ss) node [midway, yshift=6pt] {\(\hat{\vect{x}}_{\mathrm{SS}}\)};

	\draw [arrow] (x_se) -- (m_SE_env);
	\draw [arrow] (x_ss) -- (m_SS_env);

	\draw [arrow] (m_SE_env) -- (x_se_env) node [midway, yshift=7pt] {\(\tilde{\vect{x}}_{\mathrm{SE}}\)};
	\draw [arrow] (m_SS_env) -- (x_ss_env) node [midway, yshift=7pt] {\(\tilde{\vect{x}}_{\mathrm{SS}}\)};

	\draw [arrow] (x_se_env) -- (loss|-x_se_env) -- (loss);
	\draw [arrow] (x_ss_env) -- (loss|-x_ss_env) -- (loss);

	\draw [arrow] (m_filt) -- (m_SE_env);
	\draw [arrow] (m_filt) -- (m_SS_env);

	\node (h_1) [helper, at=(loss.north)] {};
	\node (h_2) [helper, at=(loss|-x_se_env)] {};
	\node (h_3) [helper, at=(x_se_env.east)] {};
	\node (h_4) [helper, at=(x_se_env.west)] {};
	\node (h_5) [helper, at=(m_SE_env.east)] {};
	\node (h_6) [helper, at=(m_SE_env.west)] {};
	\node (h_7) [helper, at=(x_se.east)] {};
	\node (h_8) [helper, at=(x_se.west)] {};
	\node (h_9) [helper, at=(m_SE.east)] {};

	\draw [line_back] (h_1.west) -- (h_2.south west) -- (h_3.south);
	\draw [line_back] (h_4.south) -- (h_5.south);
	\draw [line_back] (h_6.south) -- (h_7.south);
	\draw [arrow_back] (h_9.south) -- (h_8.south);

\end{tikzpicture}}
	\caption{Envelope regularization}
	\label{fig:env_reg}
\end{figure*}

While latent representations effectively capture semantic speech content, they often lack precise temporal information, in part due to their invariance to small time-shifts.
To address this, we propose envelope regularization to preserve the temporal structure of the enhanced output.
Our approach leverages the observation that speech dominates the signal envelope of noisy recordings.
As depicted in Figure~\ref{fig:env_reg}, we extract envelopes from both the \gls{se} output \(\hat{\vect{x}}_{\mathrm{SE}}\) and a \gls{ss} baseline \(\hat{\vect{x}}_{\mathrm{SS}}\)~\cite{speech_enhancement} using the magnitude of the Hilbert transform \(\vect{h}_{\mathrm{H}}\)~\cite{dsp_opp}.
Preliminary experiments showed that using spectral subtraction as a reference outperformed direct comparison with the noisy envelope, providing a cleaner temporal guide while remaining computationally efficient.
\par
The regularization loss is computed frame-wise as the weighted cosine similarity between these envelopes, with weights \(\vect{\rho}\) determined by the signal energy to focus on frames with speech activity
\begin{align}
	\mathcal{L}_{\mathrm{R}} = \sum_i \rho_i \cdot \operatorname{sim}(\tilde{\vect{x}}_{\mathrm{SE},i},\tilde{\vect{x}}_{\mathrm{SS},i}).
\end{align}
To compute the weight \(\rho_i\) for frame \(i\), the softmax over the frame powers of \(\hat{\vect{x}}_{\mathrm{SE}}\) is computed
\begin{align}
	\rho_i = \underset{i}{\operatorname{softmax}}\left(\tfrac{1}{\tau}\Vert\hat{\vect{x}}_{\mathrm{SE},i}\Vert^{2}\right),
\end{align}
where \(\tau \in \mathbb{R}\) represents a temperature parameter and \(\hat{\vect{x}}_{\mathrm{SE},i}\) denotes the \(i\)-th frame of the \gls{se} output.
Notably, the loss gradient is not calculated with respect to the weights \(\rho_i\).
\par
The combined LaDen loss can be written as
\begin{align}
	\mathcal{L} = \mathbb{I}_{\mathcal{L}_{\mathrm{LD}}\leq \gamma} \left(\mathcal{L}_{\mathrm{LD}} + \lambda \mathcal{L}_{\mathrm{R}} \right),
\end{align}
where \(\lambda=0.1\) is a weighting factor and \(\mathbb{I}_{\mathcal{L}_{\mathrm{LD}}\leq \gamma}\) is an indicator function that enforces an upper threshold of \(\gamma=0.05\) to the latent denoising loss.
This serves to reduce the impact of outliers and is comparable to using a threshold on the model's confidence \cite{eata}.
\par
For adaptation stability and computational efficiency, we only adapt the layer normalization and output layers of the model's parameters~\cite{tent}.

\subsection{Weight Averaging}
Inspired by ROID~\cite{roid}, a continual weight averaging is used to prevent unstable optimization and catastrophic forgetting.
After each optimization step \(t\), a linear interpolation between the adapted weights \(\theta_{t}\) and the source weights \(\theta_{\mathrm{S}}\) is performed
\begin{align}
	\theta_{t} \leftarrow \beta \theta_{t} + (1-\beta) \theta_{\mathrm{S}}.
\end{align}
This creates a favorable balance between adaptation capability and stability, allowing the model to learn from target data while maintaining the robust performance of the source model.

\section{EXPERIMENTS}
\label{sec:exp}
\subsection{Datasets}
\label{sec:datasets}
The source model is trained on the EARS datasets~\cite{ears}, containing 100h of clean speech from 107 speakers across seven speech styles (regular, loud, whisper, etc.).
Following the original publication, we utilize the EARS-WHAM dataset (denoted EARS-W), which combines EARS recordings with ambient noise from the WHAM! dataset~\cite{wham} at SNR values ranging from -2.5 to 17.5 dB for training and 0 to 20 dB for testing.
All experiments use a sampling rate of 16 kHz.
\par
To evaluate the \gls{tta} methods, we construct multiple target datasets representing different domain shifts.
In accordance with the \gls{tta} paradigm, adaptation is limited to the test-time, i.e., only the test split is used.
\subsubsection{Noise domain shift}
We create EARS-DEMAND (EARS-D) by combining clean EARS speech with noise from the DEMAND dataset~\cite{demand}, following the same mixing procedure described in \cite{ears} with SNR values ranging from -2.5 to 17.5 dB.
As the noisy environments recorded for DEMAND differ from the environments in WHAM!, this isolates adaptation to unseen background environments.
The standard test split is used for adaptation, containing 4 hours of speech from 6 speakers.
\subsubsection{Speaker and noise domain shifts}
We utilize VoiceBank+DEMAND (VBD) \cite{vbd} and create VoiceBank+WHAM (VBW) by mixing VoiceBank speech \cite{voicebank} with WHAM! noise according to the EARS mixing procedure.
For both datasets the standard test split, containing 35 minutes from two speakers, is used for adaptation.
\subsubsection{Language domain shift}
To assess the adaptation performance to unseen languages, we employ the DNS dataset~\cite{dns_2022} which contains speech in six languages (English, Russian, German, Italian, Spanish, and French) with between 18 and 90 minutes of speech per language for adaptation.
For this work, the dataset is used without additional room impulse responses.
While reverberation presents an important challenge for speech enhancement systems, we focus exclusively on additive noise scenarios in this initial exploration of \gls{tta} for \gls{se}, leaving reverberant conditions for future work.
\par
This comprehensive benchmark enables evaluation across multiple realistic domain shifts that speech enhancement systems encounter in practice.

\subsection{Models}
\begin{figure*}[h]
	\centering
	\begin{subfigure}[t]{0.4\textwidth}
		\centering
		\resizebox{\textwidth}{!}{\begin{tikzpicture}
	\tikzstyle{mlp} = [trainable, inner sep=0.75em, align=center, draw=orange, fill=orange!20, text=orange!95!black]
	\tikzstyle{residual} = [trainable, inner sep=0.75em, align=center, draw=mittelblau, fill=mittelblau!20, text=mittelblau]
	\tikzstyle{helper} = [inner sep=3pt]

	\setlength \scorner {2pt}
	\setlength \corner {5pt}

	\tikzstyle{stft} = [dsp, inner sep=0.75em, align=center]
	\tikzstyle{norm} = [trainable, inner sep=0.5em, align=center, draw=apfelgruen, fill=apfelgruen!20, text=apfelgruen!80!black]

	\node (x_in) {\(\matr{Y}\)};

	\node (input) [right=3em of x_in, mlp] {\rotatebox{90}{MLP}};
	\node (res_1) [right=1em of input, residual] {\rotatebox{90}{Residual}};
	\node (res_2) [right=2em of res_1, residual] {\rotatebox{90}{Residual}};
	\node (out_block) [right=1em of res_2, mlp] {\rotatebox{90}{MLP}};
	\node (mult) [right=2em of out_block, draw, circle, inner sep=0.0em] {\(\odot\)};
	\node (out) [right=1em of mult] {$\hat{\matr{X}}$};

	\node (bound) [draw=lila, fit=(input)(out_block)(res_1), inner sep=10pt, dashed, thick] {};
	\node (h_1) [below=0.5em of bound, helper] {};
	\node (h_2) [at=($(x_in.east)!0.5!(bound.west)$), helper] {};

	\draw [arrow] (x_in) -- (input);
	\draw [arrow] (input) -- (res_1);
	\draw [arrow, dotted] (res_1) -- (res_2) node [midway, yshift=-10pt] {$\times L$};
	\draw [arrow] (res_2) -- (out_block);
	\draw [arrow] (out_block) -- (mult);
	\draw [arrow] (mult) -- (out);
	\draw [arrow] (h_2.center) -- (h_1-|h_2) -- (h_1-|mult) -- (mult);

\end{tikzpicture}}
		\caption{Model architecture}
		\label{fig:arch_schem}
	\end{subfigure}
	\begin{subfigure}[t]{0.52\textwidth}
		\centering
		\resizebox{\textwidth}{!}{\begin{tikzpicture}
	\tikzstyle{model} = [trainable, inner sep=0.5em, align=center]
	\tikzstyle{norm} = [trainable, inner sep=0.5em, align=center, draw=apfelgruen, fill=apfelgruen!20, text=apfelgruen!80!black]
	\tikzstyle{mlp} = [trainable, inner sep=0.75em, align=center, draw=orange, fill=orange!20, text=orange!95!black]
	\tikzstyle{temp} = [trainable, inner sep=0.75em, align=center, draw=pink, fill=pink!15, text=pink!90!black]
	\tikzstyle{local} = [trainable, inner sep=0.75em, align=center, draw=lavendel, fill=lavendel!20, text=lavendel!95!black]
	\tikzstyle{helper} = [inner sep=3pt]
	\tikzstyle{adder} = [inner sep=0pt, circle, draw]

    \setlength \scorner {2pt}
    \setlength \corner {5pt}

    \node (x_in) {};

	\node (norm_1) [right=3em of x_in, norm] {\rotatebox{90}{Layer Norm}};
	\node (time) [right=2em of norm_1, temp] {\rotatebox{90}{Self-Attention}};
	\node (add_1) [right=8pt of time, adder] {\(+\)};
	\node (freq_1) [right=2em of add_1, mlp] {\rotatebox{90}{MLP}};
	\node (add_2) [right=8pt of freq_1, adder] {\(+\)};
	\node (local) [right=2em of add_2, local] {\rotatebox{90}{Conv2D}};
	\node (add_3) [right=8pt of local, adder] {\(+\)};
	\node (norm_2) [right=8pt of add_3, norm] {\rotatebox{90}{Layer Norm}};
	\node (freq_2) [right=1em of norm_2, mlp] {\rotatebox{90}{MLP}};
	\node (add_4) [right=8pt of freq_2, adder] {\(+\)};

	\node (h_1) [helper, left=1em of norm_1] {};
	\node (h_2) [helper, at=($(norm_1)!0.5!(time)$)] {};
	\node (h_3) [helper, at=($(add_1.east)!0.5!(freq_1.west)$)] {};
	\node (h_4) [helper, at=($(add_2)!0.5!(local)$)] {};
	\node (h_5) [helper, at=($(norm_1)!0.5!(time)$)] {};
	\node (h_6) [helper, above=0.3em of time] {};
	\node (h_9) [helper, above=0.5em of local] {};
	\node (h_7) [helper, above=1em of freq_1] {};
	\node (h_8) [helper, above=1em of h_7] {};

    \draw[arrow] (h_2.center) -- (h_2|-h_6) -- (add_1|-h_6) -- (add_1);
    \draw[arrow] (h_3.center) -- (h_3|-h_7) -- (add_2|-h_7) -- (add_2);
    \draw[arrow] (h_4.center) -- (h_4|-h_9) -- (add_3|-h_9) -- (add_3);
    \draw[arrow] (h_1.center) -- (h_1|-h_8) -- (add_4|-h_8) -- (add_4);

	\node (bound) [draw, dashed, fit=(h_1)(h_6)(h_8)(norm_1)(add_4)(freq_1)(time), inner xsep=0.5em, inner ysep=0.5em, mittelblau, thick] {};

	\node (x_out) [right=of add_4] {};

	\draw [arrow] (x_in) -- (norm_1);
	\draw [arrow] (norm_1) -- (time);
	\draw [arrow] (time) -- (add_1);
	\draw [arrow] (add_1) -- (freq_1);
	\draw [arrow] (freq_1) -- (add_2);
	\draw [arrow] (add_2) -- (local);
	\draw [arrow] (local) -- (add_3);
	\draw [arrow] (add_3) -- (norm_2);
	\draw [arrow] (norm_2) -- (freq_2);
	\draw [arrow] (freq_2) -- (add_4);
	\draw [arrow] (add_4) -- (x_out);

\end{tikzpicture}}
		\caption{Residual block}
		\label{fig:res_schem}
	\end{subfigure}
	\caption{Architecture of the \gls{am} model.}
	\label{fig:am_arch}
\end{figure*}
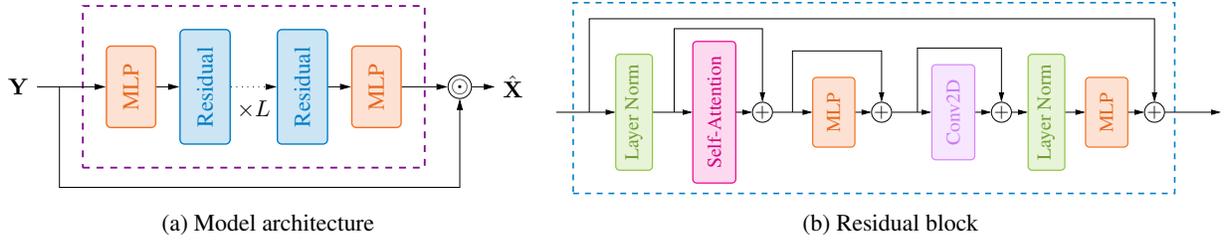

The proposed method is evaluated using two model architectures that represent a wide range of \gls{se} architectures.
The first architecture represents simple \acrfull{am} approaches common in many \gls{se} systems.
Figure~\ref{fig:arch_schem} illustrates the overall architecture of the model.
It consists of \(L\) residual blocks that sequentially transform the input STFT features to an amplitude mask.
Each residual block (detailed in~\ref{fig:res_schem}) contains MLP layers that operate along the frequency-dimension with shared weights across time steps, self-attention that applies scaled dot-product attention along the time dimension, and Conv2D layers using multi-dilated convolutions for joint time-frequency processing.
The input MLP expands the frequency dimension to 256, while the output MLP projects back to the original frequency dimension.
The enhanced magnitude is combined with the noisy phase before transforming to the time domain via an iSTFT.
The model is trained using a \gls{mse} loss that focuses on signal reconstruction quality.
In this work we used \(L=3\) residual blocks, resulting in a total of approximately 1.5M parameters, of which 123K are adapted.
In the following, this architecture is denoted as \emph{AM}.
\par
Secondly, the popular CMGAN~\cite{cmgan} is used to represent the current trend of state-of-the-art models.
It combines an encoder-decoder structure based on DenseNet~\cite{densenet} with Conformer blocks~\cite{conformer} and implicitly estimates phase components rather than just the magnitude.
Following current trends in \gls{se}, CMGAN prioritizes perceptual performance over signal level metrics through a MetricGAN~\cite{metric_gan} based loss function.
Of the 1.8M generator parameters, 6K are adapted.
Complete architectural details and hyperparameters are provided in~\cite{cmgan} and our open-source framework.

\subsection{Baselines}
To assess the proposed method, the unadapted source models and RemixIT~(\cite{remixit}, see Section~\ref{sec:related}) are used as baselines.
As RemixIT was not designed with \gls{tta} in mind, it is adjusted to work in the \gls{tta} setting.
To conform to the online setting, i.e. only one epoch with simultaneous adaptation and denoising, the teacher is updated every \(U=8\) batches instead of after each epoch.
We use an exponentially moving average teacher as proposed in~\cite{remixit} to maintain stability given the short update intervals.
Additionally, permuting speech and noise estimates is performed for each batch individually.
As recommended in~\cite{remixit}, the \gls{mse} loss is used to adapt the model on the bootstrapped dataset.

\subsection{Metrics}
All considered \gls{tta} methods are evaluated using standard \gls{se} metrics for evaluating speech quality.
As is common in \gls{se}, this work puts an emphasis on perceptual metrics.
These include \acrshort{pesq}~\cite{pesq} and the composite measures CSIG, CBAK and COVL \cite{mos}.
To also evaluate the signal level quality, the metrics \acrshort{ssnr} and \acrshort{sisdr} are used~\cite{speech_enhancement}.
As \gls{tta} requires simultaneous adaptation and inference, the average of the metrics over the adaptation period is reported.
Depending on the dataset, the adaptation period comprises between 100 and 900 utterances, each lasting 1-30s.

\subsection{Experimental Details}
\label{sec:exp_setting}
The AdamW optimizer~\cite{adam_w} is used for both source training and adaptation, with learning rates \(\alpha\) of \(1\cdot 10^{-3}\) and \(5\cdot 10^{-4}\), respectively.
Additionally, the code framework is published along with instruction on how to reproduce the results.~\footnote{Code available at: \url{https://github.com/tobiaaa/SETTA}}
All experiments were conducted using a single Nvidia A6000 GPU.
The central experiments are repeated 10 times to assess the statistical significance of the results.
Since \gls{tta} does not depend on random model initializations, the main cause of randomness is the order of the data, which varies between experiments.

\section{RESULTS}

\subsection{Result Analysis}
Table~\ref{tab:avg_comb} shows the performance of the AM architecture averaged across the target datasets.
As LaDen's adaptation objective is based on WavLM embeddings, which contain high-level, perceptual information, the adapted model performs better on the perceptual metrics than on the signal level metrics.
This also explains the relatively small gain on the CBAK metric.
WavLM's encoder, trained with a HuBERT-like masked prediction loss~\cite{hubert}, prioritizes distinguishing between time frames.
Since background noise is typically more stationary than speech, it receives less representation in the embeddings, resulting in adaptation that focuses less on the background.
\begin{table*}[h]
	\def\arraystretch{1}
	\centering
	\caption{Results for both architectures averaged over the datasets (\(\mu\pm2\sigma\)). SSNR and SI-SDR in dB.}
	\label{tab:avg_comb}
	\begin{tabular}{llcccccc}
	\toprule
	                                        &         & \avg{PESQ} \(\uparrow\)  & \avg{CSIG} \(\uparrow\)  & \avg{CBAK} \(\uparrow\)  & \avg{COVL} \(\uparrow\)  & \avg{SSNR} \(\uparrow\) & \avg{SI-SDR} \(\uparrow\) \\
	\midrule
	\multirow{3}{\wcol}{\rotatebox{90}{AM}} & Source  & 2.05                     & 3.07                     & 2.78                     & 2.52                     & 7.42                    & 12.28                     \\
	                                        & RemixIT & 2.06\(\pm\).006          & 3.10\(\pm\).007          & \textbf{2.80}\(\pm\).005 & 2.54\(\pm\).006          & \textbf{7.48}\(\pm\).03 & \textbf{12.47}\(\pm\).04  \\
	                                        & LaDen   & \textbf{2.13}\(\pm\).005 & \textbf{3.13}\(\pm\).007 & \textbf{2.80}\(\pm\).005 & \textbf{2.59}\(\pm\).006 & 7.01\(\pm\).04          & 12.33\(\pm\).04           \\
	\midrule
	\multirow{3}{\wcol}{\rotatebox{90}{CMGAN}} & Source  & 2.60                     & 3.75                     & 3.02                     & 3.15                     & 6.00                    & 11.32                     \\
	                                        & RemixIT & 2.60\(\pm\).006          & 3.77\(\pm\).006          & 3.03\(\pm\).007          & 3.17\(\pm\).007          & 5.92\(\pm\).07          & 11.52\(\pm\).07           \\
	                                        & LaDen   & \textbf{2.62}\(\pm\).002 & \textbf{3.81}\(\pm\).002 & \textbf{3.07}\(\pm\).002 & \textbf{3.20}\(\pm\).002 & \textbf{6.31}\(\pm\).03 & \textbf{12.09}\(\pm\).02  \\
	\bottomrule
\end{tabular}

\end{table*}
\par
\begin{figure}[H]
	\def \del {\(\Delta\)}
	\centering
	\begin{subfigure}[b]{\columnwidth}
		\centering
		\resizebox{0.8\textwidth}{!}{\begin{tikzpicture}
	\tkzKiviatDiagram[scale=0.9,label distance=0.0cm,
		gap=1.0,
		ymin=-0.15,
		ymax=0.15,
		label space=4.2cm,
		lattice=3]{EARS-D,VBD,VBW,\dns{EN},\dns{GE},\dns{IT},\dns{RU},\dns{SP},\dns{FR}}
	\tkzKiviatLine[thick,color=mittelblau,mark=none,opacity=.5](
	0.0,
	0.0,
	0.0,
	0.0,
	0.0,
	0.0,
	0.0,
	0.0,
	0.0
	)
	\tkzKiviatLineError[thick,color=orange,opacity=.5](
	-0.056/0.023,
	-0.021/0.010,
	0.003/0.009,
	0.032/0.020,
	0.054/0.023,
	0.036/0.018,
	0.004/0.014,
	0.029/0.024,
	0.037/0.017
	)
	\tkzKiviatLineError[thick,mark size=4pt,color =lila](
	-0.053/0.006,
	0.130/0.006,
	0.027/0.011,
	0.097/0.010,
	0.129/0.016,
	0.083/0.013,
	0.048/0.007,
	0.112/0.013,
	0.130/0.008
	)
	\tkzKiviatGrad[prefix=,unity=1,suffix=](0)
\end{tikzpicture}}
		\caption{\del PESQ \(\uparrow\)}
		\label{fig:res_pesq}
	\end{subfigure}
	\hfil
	\begin{subfigure}[b]{\columnwidth}
		\centering
		\resizebox{0.8\textwidth}{!}{\begin{tikzpicture}
	\tkzKiviatDiagram[scale=0.7,label distance=1.0cm,
		gap=1.0,
		ymin=-2.0,
		ymax=2.0,
		label space=5.2cm,
		lattice=4]{EARS-D,VBD,VBW,\dns{EN},\dns{GE},\dns{IT},\dns{RU},\dns{SP},\dns{FR}}
	\tkzKiviatLine[thick,color=mittelblau,mark=none,opacity=.5](
	0.0,
	0.0,
	0.0,
	0.0,
	0.0,
	0.0,
	0.0,
	0.0,
	0.0
	)
	\tkzKiviatLineError[thick,color=orange,opacity=.5](
	-0.624/0.157,
	-0.099/0.095,
	0.319/0.029,
	0.383/0.045,
	0.426/0.116,
	0.358/0.137,
	0.234/0.098,
	0.332/0.158,
	0.366/0.051
	)
	\tkzKiviatLineError[thick,mark size=4pt,color=lila](
	0.216/0.103,
	1.494/0.107,
	-0.107/0.030,
	-0.097/0.073,
	0.040/0.077,
	-0.635/0.119,
	-0.205/0.120,
	0.220/0.097,
	-0.437/0.060
	)
	\tkzKiviatGrad[prefix=,unity=1,suffix=](0)
\end{tikzpicture}}
		\caption{\del SI-SDR [dB] \(\uparrow\)}
		\label{fig:res_sisdr}
	\end{subfigure}
	\begin{subfigure}[b]{\columnwidth}
		\centering
		\resizebox{0.7\textwidth}{!}{\begin{tikzpicture}
    \begin{axis}[%
            hide axis,
            xmin=10,
            xmax=50,
            ymin=0,
            ymax=0.4,
            legend style={font=\small, legend columns=-1},
            legend cell align={left},
        ]
        \addlegendimage{mittelblau, thick}
        \addlegendentry{Source \ }
        \addlegendimage{orange, thick}
        \addlegendentry{RemixIT \ }
        \addlegendimage{lila, thick}
        \addlegendentry{LaDen}
    \end{axis}
\end{tikzpicture}}
	\end{subfigure}
	\caption{TTA results relative to the source performance using the AM model (\(\mu\pm2\sigma\)).}
	\label{fig:res_percept}
\end{figure}
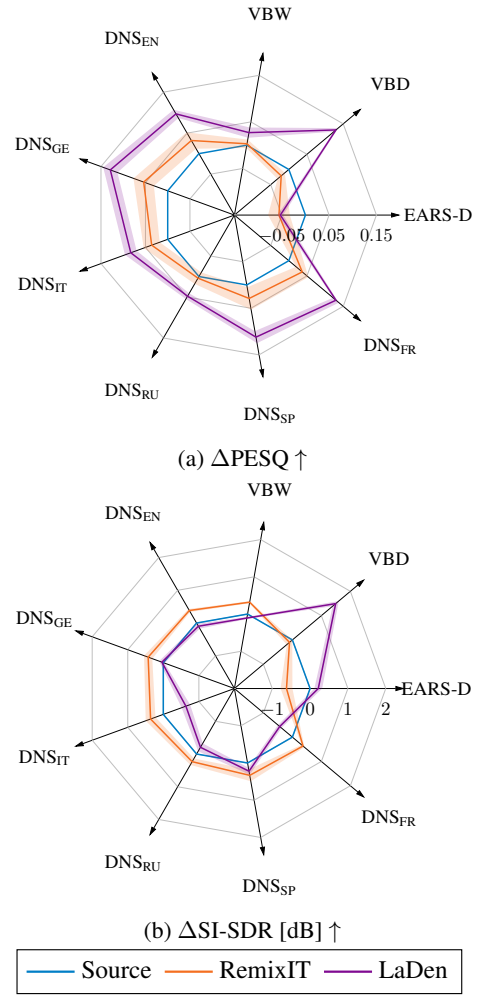

To assess the performance in more detail, Figure~\ref{fig:res_percept} shows the performance per dataset for PESQ and \gls{sisdr}.
The metrics are displayed as the difference to the source performance.
With the exception of the EARS-D dataset, LaDen achieves a significantly larger perceptual improvement over the source performance than RemixIT.
On the EARS-D dataset, neither of the \gls{tta} methods outperforms the source model.
In the context of the remarkable accuracy of \gls{diet} on this dataset (cf. Table~\ref{tab:mat_gen}), the proposed pseudo-labeling is likely not the main limiting factor.
This suggests that the source model generalizes well for shifts only in the noise distribution, leaving little room for improvement through adaptation (cf. Table~\ref{tab:pre_comp_cmgan}).
In case of speaker and language domain shifts, the source model does not generalize as well and \gls{laden} demonstrates significant and consistent improvements.
On the signal level metrics, RemixIT achieves a more consistent gain compared to LaDen.
As an exception, LaDen achieves outstanding results across all metrics on the \gls{vbd} datasets using the AM architecture.
While the exact reasons for this pattern require further investigation, it suggests that the AM model has significant room for improvement on speaker and noise domain shifts that LaDen is able to fill.
\par
The results of \gls{tta} for \gls{se} using the CMGAN architecture are listed in Table~\ref{tab:avg_comb}.
Notably, the baseline source performance reflects the perceptual focus of MetricGAN used in CMGAN.
Interestingly, in this setting, LaDen achieves a significant gain on the signal level metrics without diminishing the outstanding perceptual performance.
In contrast, RemixIT is not able to substantially improve upon the source performance on any of the metrics.
\par
\begin{figure}[h]
	\def \del {\(\Delta\)}
	\centering
	\begin{subfigure}[b]{\columnwidth}
		\centering
		\resizebox{0.8\textwidth}{!}{\begin{tikzpicture}
	\tkzKiviatDiagram[scale=0.9,label distance=0.0cm,
		gap=1.0,
		ymin=-0.2,
		ymax=0.1,
		label space=4.2cm,
		lattice=3]{EARS-D,VBD,VBW,\dns{EN},\dns{GE},\dns{IT},\dns{RU},\dns{SP},\dns{FR}}
	\tkzKiviatLine[thick,color=mittelblau,mark=none,opacity=.5](
	0.0,
	0.0,
	0.0,
	0.0,
	0.0,
	0.0,
	0.0,
	0.0,
	0.0
	)
	\tkzKiviatLineError[thick,color=orange,opacity=.5](
	-0.007/0.006,
	-0.018/0.003,
	0.018/0.007,
	0.004/0.007,
	0.012/0.018,
	0.003/0.029,
	-0.001/0.005,
	0.005/0.007,
	0.009/0.030
	)
	\tkzKiviatLineError[thick,mark size=4pt,color =lila](
	-0.056/0.007,
	-0.081/0.006,
	-0.010/0.003,
	0.016/0.005,
	0.075/0.010,
	0.091/0.009,
	-0.002/0.004,
	0.046/0.006,
	0.102/0.012
	)
	\tkzKiviatGrad[prefix=,unity=1,suffix=](0)
\end{tikzpicture}}
		\caption{\del PESQ \(\uparrow\)}
		\label{fig:res_pesq_cmgan}
	\end{subfigure}
	\hfil
	\begin{subfigure}[b]{\columnwidth}
		\centering
		\resizebox{0.8\textwidth}{!}{\begin{tikzpicture}
	\tkzKiviatDiagram[scale=0.8,label distance=1.0cm,
		gap=1.0,
		ymin=-2.0,
		ymax=2.0,
		label space=5.2cm,
		lattice=4]{EARS-D,VBD,VBW,\dns{EN},\dns{GE},\dns{IT},\dns{RU},\dns{SP},\dns{FR}}
	\tkzKiviatLine[thick,color=mittelblau,mark=none,opacity=.5](
	0.0,
	0.0,
	0.0,
	0.0,
	0.0,
	0.0,
	0.0,
	0.0,
	0.0
	)
	\tkzKiviatLineError[thick,color=orange,opacity=.5](
	0.026/0.083,
	0.291/0.068,
	1.135/0.314,
	-0.003/0.155,
	0.087/0.287,
	0.100/0.218,
	0.049/0.049,
	-0.006/0.103,
	0.125/0.229
	)
	\tkzKiviatLineError[thick,mark size=4pt,color =lila](
	0.453/0.022,
	0.945/0.029,
	1.683/0.026,
	0.555/0.040,
	1.026/0.048,
	0.568/0.114,
	0.058/0.021,
	0.629/0.045,
	1.008/0.094
	)
	\tkzKiviatGrad[prefix=,unity=1,suffix=](0)
\end{tikzpicture}}
		\caption{\del SI-SDR [dB] \(\uparrow\)}
		\label{fig:res_sisdr_cmgan}
	\end{subfigure}
	\begin{subfigure}[b]{\columnwidth}
		\centering
		\vspace{5pt}
		\resizebox{0.7\textwidth}{!}{\begin{tikzpicture}
    \begin{axis}[%
            hide axis,
            xmin=10,
            xmax=50,
            ymin=0,
            ymax=0.4,
            legend style={font=\small, legend columns=-1},
            legend cell align={left},
        ]
        \addlegendimage{mittelblau, thick}
        \addlegendentry{Source \ }
        \addlegendimage{orange, thick}
        \addlegendentry{RemixIT \ }
        \addlegendimage{lila, thick}
        \addlegendentry{LaDen}
    \end{axis}
\end{tikzpicture}}
	\end{subfigure}
	\caption{TTA results relative to the source performance using CMGAN (\(\mu\pm2\sigma\)).}
	\label{fig:res_cmgan}
\end{figure}
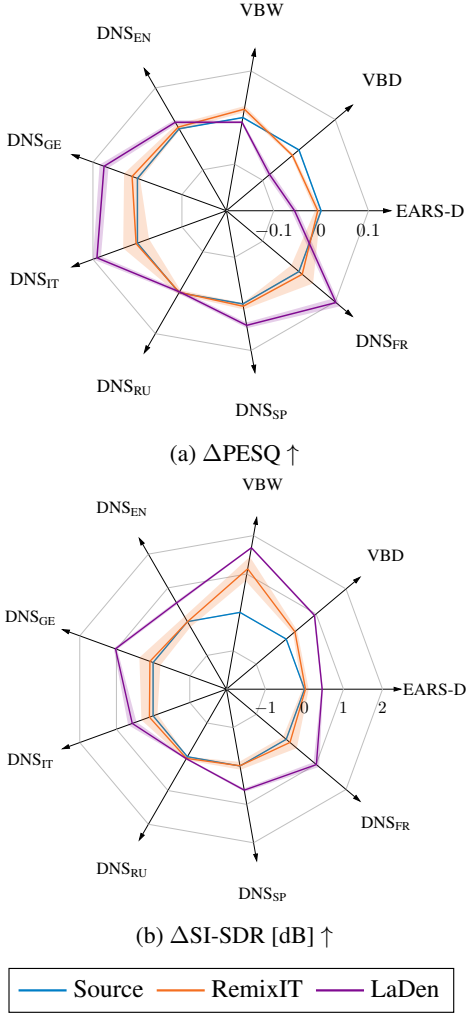

Examining the dataset specific perceptual results depicted in Figure~\ref{fig:res_pesq_cmgan}, RemixIT closely adheres to the source performance, whereas LaDen's results are more mixed, achieving a significant gain on the DNS-based datasets, at the cost of diminished performance on the EARS-D and VoiceBank-based datasets.
The contrasting behavior on the VBD dataset using the two architectures suggests that the amenability to adaptation depends on the underlying model architecture and source training.
On the signal level performance however (cf. Figure~\ref{fig:res_sisdr_cmgan}), LaDen achieves a consistent gain over the source model and RemixIT.
Interestingly, CMGAN exhibits a similar pattern for the EARS-D dataset as the AM architecture, confirming that no significant perceptual gain can be achieved in noise-only domain shifts.
\par
These results highlight LaDen's versatility as a \gls{tta} method for speech enhancement.
Particularly noteworthy is LaDen's ability to enhance CMGAN's signal level performance without compromising its perceptual quality.
CMGAN's MetricGAN loss puts a strong emphasis on perceptual performance, leaving little room for improvement.
Conversely, the \gls{mse} loss of the AM architecture prioritizes signal reconstruction.
In both cases, LaDen is able to complement the strengths of the trained source model by improving upon their respective shortcomings.

\subsection{Ablation Study}
Table~\ref{tab:ablation} presents the incremental impact of each component in our proposed method across both perceptual and signal level metrics.
The basic latent denoising approach shows notable improvements in perceptual quality across both datasets, but exhibits mixed results for signal level metrics.
Adding envelope regularization addresses this limitation by enforcing temporal structure.
For VBD it provides substantial improvements in both perceptual and signal level metrics.
However, for EARS-D, we observe a decrease in performance, likely due to the prevalence of silent segments in this dataset where the regularization introduces artifacts, which is mitigated via the proposed power weighting.
The final addition of weight averaging stabilizes the adaptation process, preventing performance degradation over time.
Evidently, no single configuration is ideal across all datasets and metrics, necessitating a balanced approach when adapting to unknown domain shifts.
\begin{table}[h]
	\def\arraystretch{1}
	\centering
	\caption{Ablation study on the AM architecture. \(\vect{\rho}\) and EMA represent the power weights and weight averaging introduced in Section~\ref{sec:method}, respectively. \Gls{sisdr} in dB.}
	\label{tab:ablation}
	\begin{tabular}{lcccccc}
	\toprule
	                               & \multicolumn{2}{c}{EARS-D} & \multicolumn{2}{c}{VBD} &                                         \\
	\cmidrule(r){2-3} \cmidrule(r){4-5}\cmidrule(r){6-7}
	                               & PESQ \(\uparrow\)          & SI-SDR \(\uparrow\)     & PESQ \(\uparrow\) & SI-SDR \(\uparrow\) \\
	\midrule
	Source                         & 1.974                      & 5.102                   & 2.424             & 11.487              \\
	\midrule{}
	\(\mathcal{L}_{\mathrm{LD}}\)  & \textbf{2.038}             & \textbf{6.986}          & 2.614             & 11.175              \\
	+ \(\mathcal{L}_{\mathrm{R}}\) & 1.864                      & 4.782                   & \textbf{2.642}    & 12.432              \\
	+ \(\vect{\rho}\)              & 1.887                      & 5.064                   & 2.543             & \textbf{13.097}     \\
	+ EMA                          & 2.033                      & 6.558                   & 2.591             & 12.000              \\
	\bottomrule
\end{tabular}

\end{table}

\section{DISCUSSION}
The \gls{tta} results demonstrate that LaDen provides effective adaptation for speech enhancement across multiple domain shifts and model architectures.
While neither \gls{tta} method succeeds with noise-only domain shifts, LaDen consistently outperforms RemixIT on perceptual metrics, particularly for domain shifts involving speakers or languages.
Furthermore, LaDen's ability to improve CMGAN's signal level performance without compromising its perceptual quality highlights the complementary nature of latent denoising to both the \gls{mse} loss of the AM model, as well as the perceptual approach of CMGAN.
\par
The effectiveness of \gls{diet} suggests that while domain shifts may be complex at the signal level, they become more manageable in the latent space.
Besides \gls{diet}'s impressive estimation accuracy, its ability to estimate reliable pseudo-labels further validates the underlying principle.
There are multiple reasons for a simple, even linear relationship between the embeddings of clean speech and noisy speech.
Previous studies found that large encoders semantically disentangle the structure of their input space~\cite{lin_hyp}.
In the context of natural language processing, this means semantic concepts are represented by directions in the embedding space, where causally separable concepts are represented by orthogonal vectors~\cite{lin_hyp}.
The proposed \gls{diet} translates this to the independent concepts of speech and noise in the embedding space of speech encoders.
This phenomenon is also known in vision where nonlinear transformations (e.g., lighting, composition) can be linearized in learned embeddings~\cite{capsule}.
\par
Our approach reveals fundamental differences between \gls{tta} for \gls{se} and classification tasks.
Unlike classification, where entropy serves as a natural adaptation signal and confidence heuristic, speech enhancement requires more sophisticated proxies for adaptation quality.
Furthermore, whereas the accuracy suffices in comparing classification \gls{tta} methods, \gls{se} \gls{tta} methods cannot be judged solely on their effectiveness, but also on the alignment of their adaptation objective to the task at hand, e.g. the trade-off between perceptual and signal level performance.
\par
Future research should address several promising directions.
Extending LaDen to handle reverberant conditions represents an important next step, possibly requiring specialized latent representations that capture room acoustics.
Improving adaptation for noise-only shifts, despite their currently limited gains, could benefit scenarios with highly non-stationary noise.
Finally, the extent to which \gls{diet} is applicable to other domains such as image-to-image transformation in computer vision or medical image enhancement like MRI artifact removal presents a compelling challenge for future research.
\section{CONCLUSION}
We presented LaDen, the first test-time adaptation method specifically designed for speech enhancement.
By leveraging speech representations from an existing speech encoder and performing latent denoising through a \acrlong{diet} of noisy embeddings, our approach enables effective adaptation across multiple domain shifts (noise, speaker, language) without requiring labeled target data.
Our comprehensive evaluation demonstrated LaDen's ability to improve perceptual quality across varied acoustic conditions, with particular effectiveness for speaker and language domain shifts.
LaDen's consistent performance across different model architectures and training objectives highlights its versatility as a practical solution for real-world speech enhancement systems that must adapt to previously unseen environments.
This work establishes a foundation for future research on test-time adaptation methods specifically designed for generative audio tasks.


\bibliographystyle{IEEEbib}
\bibliography{refs}

\end{document}